\documentstyle[preprint,aps]{revtex}
\begin{document}
\draft
\tightenlines
\title
{\bf Functional integral over velocities for a spinning particle
 with and without anomalous magnetic 
moment in a constant electromagnetic field }

\author
{WELLINGTON DA CRUZ\footnote{E-mail: wdacruz@fisica.uel.br}}


\address
{Departamento de F\'{\i}sica,\\
 Universidade Estadual de Londrina, Caixa Postal 6001,\\
Cep 86051-970 Londrina, PR, Brazil\\}
 
\date{\today}

\maketitle
\begin{abstract}
The technique of functional integration over velocities
 is applied
 to calculation of
 the propagator of a spinning particle with and without
  anomalous
  magnetic 
 moment. A representation for the spin factor is obtained
  in this
  context for the 
 particle in a constant electromagnetic field. As a
  by-product, we 
also obtained a Schwinger representation for the
 first case.
\end{abstract}
\pacs{PACS numbers: 11.10.Ef; 03.65.Pm \\
Keywords: Path integrals; Grassmann variables; Spin factor; 
Schwinger \\representation }

\newpage

Functional integration over velocities has been implemented
 to calculate 
the
causal propagator of a relativistic spinless particle\cite{R1}
 and we have 
 established some rules for handling the Gaussian and
 quasi-Gaussian integrals. 
In our representation the integration over velocities
 does not have any 
restrictions imposed by boundary conditions and the
 matrices
 obtained after integration
 do
not contain any derivatives in time. On the other hand,
 Gitman and Shvartsman
\cite{R2} have obtained for the spinning particle  
{\it \`a la } Polyakov
\cite{R3} a 
bosonic functional representation with a spin factor
 where the spinor 
structure 
of the integrand is written as a decomposition of 
independent $\gamma$-matrix.
In this paper we obtain the propagator
 for a spinning 
particle with and without anomalous magnetic moment. In the 
course 
of calculation we show that the spin factor for the
 particle in a 
 constant electromagnetic field can be obtained
  straightforwardly 
 because in this case it does not depend on the
  trajectories. 
 Our results coincide with others by Gitman 
 {\it et al}\cite{R4}. 
 The paper is organized as follows: first we
  consider the
  spinning 
 particle without anomalous magnetic moment in much
 more detail and for the case with anomalous
  magnetic moment 
 we skip some steps that would otherwise be
  repeated. 
 However, important matrix functions that
  appear in the 
 equations are given below for reference. Our
  notation is the same 
 as in the cited papers.
 The propagator of the spinning particle has
  the following
  form\cite{R5}
  :

 \begin{eqnarray}
 \label{e1}
&& {\tilde S}^c(x_{out},x_{in})=\exp\left(i\gamma^{n}
\frac{\partial_{l}}
 {\partial \theta^n}\right)\int_{0}^{\infty}\;de_{0}
 \;\int\; d\chi_{0}
 \\
&&\times\int \exp\left\{ i\int_{0}^1\left[-\frac
{\dot{x}^2}{2e}
-\frac{em^2}{2}-g{\dot x}A(x)+iegF_{\mu\nu}(x)\psi^{\mu}
\psi^{\nu}
\right.\right.\nonumber\\
&&+\left.\left.i\left(\frac{{\dot x}_{\mu}\psi^{\mu}}{e}
-m\psi^5\right)\chi
-i\psi_{n}{\dot{\psi}^{n}}+\pi{\dot{e}}+\nu{\dot {\chi}}
\right]d\tau+
\psi_{n}(1)\psi^n(0)\right\}\nonumber\\
&&\times\left. M(e)\; Dx\; De\; D\pi\; D\chi\; D\nu
\;{\cal D}\psi\;
 \right|_{\theta=0},\nonumber
 \end{eqnarray}
 
 \noindent where
 
 \[
 M(e)=\int Dp\; \exp \left\{\frac{i}{2}\int_{0}^1 e p^2
 \; d\tau\right\},\\
 {\cal D\psi}=D\psi\left[\int_{\psi(1)+\psi(0)=\theta}D\psi
 \exp\left\{\int_{0}^1\psi_{n}{\dot\psi}^n d\tau\right\}
 \right]^{-1},\\
 \]
 \noindent and
 \[
 {\tilde S}^c=S^c\gamma^5.
 \]
 
 \noindent Performing integration over $\pi$ and $\nu$
  we get
 
 \[
 \int De\;\delta({\dot {e}})f(e)=f(e_{0}),\;\int D\chi\; 
 \delta(\dot{\chi})f(\chi)=f(\chi_{0}).
 \]
 So,
 
 \begin{eqnarray}
 &&{\tilde S}^c(x_{out},x_{in})=\exp\left(i\gamma^n
 \frac{\partial_{l}}{\partial\theta^n}\right)\int_{0}^{\infty}
 de_{0}\;Dp\;d\chi_{0}\;Dx\;{\cal D}\psi\\
 &&\times\exp\left\{i\int_{0}^1\left[-\frac{{\dot x}^2}{2e_{0}}
 -\frac{e_{0}m^2}{2}-g{\dot x}A+\frac{e_{0}p^2}{2}\right]d\tau
 \right\}\nonumber\\
 &&\times\exp\left\{i\int_{0}^1\left[ige_{0}F_{\alpha\beta}
 \psi^{\alpha}\psi^{\beta}+i\left(\frac{{\dot x}_{\alpha}
 \psi^{\alpha}}
 {e_{0}}-m\psi^5\right)\chi_{0}-i\psi_{n}{\dot \psi}^n\right]
 d\tau\right.\nonumber\\
 &&\left.\left.+\psi_{n}(1)\psi^{n}(0)\right\}
 \right|_{\theta=0},\nonumber
 \end{eqnarray}
 
 \noindent and making the replacement
 
 \[
 {\sqrt {e}}p\rightarrow p, \; 
 \frac{x-x_{in}-\tau{\Delta x}}{\sqrt e} \rightarrow x,\;
 {\Delta x}=x_{out}-x_{in},\\
 \]
 \noindent we have new boundary conditions $x(0)=0=x(1)$.
 
 Now consider,
 
 \begin{eqnarray}
 &&\Delta (e_{0})=\left.\int_{0}^{\infty}\frac{de_{0}}
 {e_{0}^2}
 \exp\left[-\frac{i}{2}\left(e_{0}m^2+\frac{{\Delta x^2}}
 {e_{0}}\right)\right]\int_{0}^{0}Dx \int Dp\; \delta^4
 \left(\int v d\tau
 \right)
 \right.\nonumber\\
 &&\times\exp\left\{i\int d\tau \left[-\frac{{\dot x}^2}{2}
 +\frac{p^2}{2}-g\left(\sqrt{e_{0}}{\dot x}+{\Delta x}\right)
 A \left({\sqrt e_{0}}x+x_{in}
 +\tau{\Delta x}\right)\right]\right\}\nonumber
 \end{eqnarray}
 
 \noindent and we replace in this step the integrations
  over the
 trajectories by ones over velocities
 
 \[
 x(\tau)=\int_{0}^1\theta(\tau-\tau^{\prime})v
 (\tau^{\prime})d\tau^{\prime}=
 \int_{0}^{\tau}v(\tau^{\prime})d\tau^{\prime}
 ,\;\; v(\tau)=
 {\dot x}(\tau),\;\; J= {\rm Det}\;
 \theta(\tau-\tau^{\prime})
 \]
 \noindent where J is the Jacobian of transformation.
 
 Then,
 
 \begin{eqnarray}
 &&\Delta(e_{0})=\int_{0}^{\infty}\frac{de_{0}}{e_{0}^2}
 \exp\left[-\frac{i}{2}\left(e_{0}m^2+\frac{{\Delta x}^2}
 {e_{0}}
 \right)\right]\int Dv\; J \int Dp\; \delta^4
 \left(\int v d\tau \right)\\
 &&\times\exp\left\{i\int d\tau\left[-\frac{v^2}{2}
 +\frac{p^2}
 {2}-g(\sqrt{e_{0}}v+\Delta x)A \left(\sqrt{e_{0}}
 \int_{0}^{\tau} v(\tau^{\prime})
 d\tau^{\prime}+x_{in}+\tau\Delta x\right)\right]
 \right\}.\nonumber
 \end{eqnarray}
 
 \noindent On the other hand, we know that
 
 \[
 J=\frac{1}{i(2\pi)^2}\left[\int Dv\; Dp\; \delta^4
 \left(\int v 
 d\tau \right)\exp\left\{i\int d \tau\left
 (-\frac{v^2}{2}+
 \frac{p^2}{2}\right)\right\}\right]^{-1}
 \]
 \noindent and
 
 \begin{eqnarray}
 &&\Delta (e_{0})=\frac{1}{i(2\pi)^2}\int\frac{de_{0}}
 {e_{0}^2}
 \exp\left[-\frac{i}{2}\left(e_{0}m^2+\frac{{\Delta x}^2}
 {e_{0}}
 \right)\right]\int {\cal D}v\; \delta^4\left
 (\int v d\tau \right)
 \\
 &&\times\exp\left\{i\int d\tau \left[-\frac{v^2}{2}
 -g(\sqrt{e_{0}}v+\Delta x)A \left({\sqrt e_{0}}
 \int_{0}^{\tau} v(\tau^{\prime})
 d\tau^{\prime}+x_{in}+\tau\Delta x\right)\right]\right\}
 \;,\nonumber
 \end{eqnarray}
 
 \noindent where the new measure ${\cal D}v$ has the form
 
 \[
 {\cal D}v=Dv\left[\int Dv\; \delta^4\left(\int v
  d\tau\right)
 \exp\left\{-i\int \frac{v^2}{2}d\tau\right\}\right]^{-1}.
 \]
 
 \noindent The integration over $\chi_{0}$ gives us for 
 the propagator
 
 \begin{eqnarray}
 &&{\tilde S}^c(x_{out},x_{in})=-\exp\left(i\gamma^n
 \frac{\partial_{l}}{\partial\theta^n}\right)\int
  \Delta(e_{0})
 {\cal D}\psi \int_{0}^1\left[\frac{v_{\alpha}
 \psi^{\alpha}}
 {\sqrt{e_{0}}}-m\psi^5\right]d\tau\nonumber\\
 &&\times\left.\exp\left\{i\int_{0}^1d\tau
 \left[ige_{0}F_{\alpha\beta}
 \psi^{\alpha}\psi^{\beta}-i\psi_{n}{\dot \psi}^{n}
 \right]+\psi_{n}(1)
 \psi^n(0)\right\}\right|_{\theta=0}.
 \end{eqnarray}
 
 \noindent Now we replace the integration over 
 $\psi$ by one over odd velocities $\omega$
 
 \[
 \psi(\tau^{\prime\prime})=\frac{1}{2}\int
 \epsilon(\tau^{\prime\prime}-\tau)\omega(\tau)d\tau+\frac
 {\theta}{2}\;,\; \omega={\dot \psi}\;,
 \]
 
 \noindent such that there are no restrictions on
  $\omega$
 and the boundary conditions for $\psi$ are satisfied.
 
 So, the propagation function takes the following form,
 
 \begin{eqnarray}
 &&{\tilde S}^c(x_{out},x_{in})=-\exp\left(i\gamma^n
 \frac{\partial_{l}}{\partial\theta^n}\right)\int 
 \Delta(e_{0}){\cal D}\omega\\
 &&\times\int_{0}^1
 d\tau^{\prime\prime}\left[\frac{v_{\mu}}{\sqrt{e_{0}}}
 \left(\int
 \epsilon(\tau^{\prime\prime}-\tau)\omega^{\mu}(\tau)d\tau+
 \theta^{\mu}\right)-m\left(\int\epsilon(\tau^{\prime\prime}
 -\tau)\omega^5(\tau)d\tau+\theta^5\right)\right]\nonumber\\
 &&\left.\times\exp\left\{-\frac{1}{2}\int\omega^n(\tau)
 \Lambda_{nm}
 (\tau,\tau^{\prime})\omega^m(\tau^{\prime})d\tau 
 d\tau^{\prime}
 +\int Q_{n}\omega^n d\tau-\frac{1}{4}ge_{0}F_{\mu\nu}
 \theta^{\mu}\theta^{\nu}\right\}\right|_{\theta=0}
 \nonumber\;,
 \end{eqnarray}
 
 \noindent where
 
 \begin{eqnarray}
 &&\Lambda_{\mu\nu}=\epsilon(\tau-\tau^{\prime})\eta_{\mu\nu}
 -\frac{1}{2}ge_{0}
 \epsilon
 F_{\mu\nu}\epsilon\;,\nonumber\\
 &&\Lambda_{55}=\epsilon(\tau-\tau^{\prime})\eta_{55}
 \;,\; \eta_{55}=-1\;,\\
 &&Q_{5}=0\;,\; Q_{\mu}=-\frac{1}{2}ge_{0}\epsilon
  F_{\nu\mu}\theta^{\nu}\;,
 \nonumber\\
 &&{\cal D}\omega=\frac{D\omega}{\int D\omega 
 e^{-\frac{1}{2}
 \omega^n\epsilon
 \omega_{n}}}.\nonumber
 \end{eqnarray}
 
 \noindent Introducing the odd sources $\rho$ for $\omega$
  one gets
 
 \begin{eqnarray}
 &&{\tilde S}^c(x_{out},x_{in})=-\exp\left(i\gamma^n
 \frac{\partial_{l}}
 {\partial\theta^n}\right)\nonumber\\
 &&\times\int \Delta(e_{0}){\cal D}\omega\left
 [\frac{v_{\mu}}{\sqrt{e_{0}}}
 \left(\epsilon\frac{\delta}{\delta\rho_{\mu}}
 +\theta^{\mu}\right)
 -m\left(\epsilon\frac{\delta}{\delta\rho_{5}}
 +\theta^{5}\right)\right]
 \nonumber\\
 &&\left.\times\exp\left\{-\frac{1}{2}\int\omega^n 
 \Lambda_{nm}\omega^m d\tau 
 d\tau^{\prime}+\int (Q_{n}+\rho_{n})\omega^{n} d\tau
  \right\}
 \right|_{\rho=0}
 \;,
 \end{eqnarray}
 
 \noindent and we define
 
 \begin{eqnarray}
 &&I_{n}=\rho_{n}+Q_{n}\;,\nonumber\\
 &&I_{5}=\rho_{5}\;,\\
 &&I_{\mu}=\rho_{\mu}+\frac{1}{2}ge_{0}F_{\mu\nu}
 \theta^{\nu}
 \epsilon.\nonumber
 \end{eqnarray}
 
 \noindent Given that for a function in the Grassmann
  algebra
  the following 
 identity holds,
 
 \begin{eqnarray}
 \left.\exp\left(i\gamma^n\frac{\partial_{l}}{\partial
 \theta^n}
 \right)f(\theta)
 \right|_{\theta=0}&=&\left. f\left(\frac{\partial_{l}}
 {\partial
 \zeta}\right)
 \exp(i\zeta_{n}\gamma^n)\right|_{\zeta=0}\\
 &=&\left.\sum_{k=0}^{4}\sum_{n_{1}...n_{k}}f_{n_{1}...n_{k}}
 \frac{\partial_{l}}{\partial
 \zeta_{n_{1}}}...\frac{\partial_{l}}{\partial\zeta_{n_{k}}}
 \sum_{l=0}^4
 \frac{i^{l}}{l!}(\zeta_{n}\gamma^n)^{l}\right|_{\zeta=0}
 \;,\nonumber
 \end{eqnarray}
 
 \noindent where the $\zeta_{n}$ are odd variables, we find
 
 \begin{eqnarray}
 &&{\tilde S}^c(x_{out},x_{in})=-\int {\cal D}\omega\;
 {\Delta}(e_{0})\\
 &&\times\left[\frac{v_{\mu}}{\sqrt{e_{0}}}\left(\epsilon
 \frac{\delta}{\delta
 \rho_{\mu}}+\frac{\partial}{\partial\zeta_{\mu}}\right)
 -m\left(\epsilon
 \frac{\delta}{\delta\rho_{5}}+i\gamma^5\right)
 \right]\nonumber\\
 &&\times\exp\left\{-\frac{1}{4}ge_{0}F_{\mu\nu}
 \frac{\partial}
 {\partial\zeta_{\mu}}\frac{\partial}{\partial
 \zeta_{\nu}}\right\}\nonumber\\
 &&\left.\times\exp\left\{-\frac{1}{2}\omega^n \Lambda_{nm}
 \omega^{m}+I_{n}\omega^{n}
 \right\}\exp(i\zeta_{\lambda}\gamma^{\lambda})
 \right|_{\zeta=0,\rho=0}.
 \nonumber
 \end{eqnarray}
 
 \noindent After the usual shift on $\omega$, we
  perform the integration 
 that results in
 
 \begin{eqnarray}
 &&{\tilde S}^c(x_{out},x_{in})=-\frac{1}{2i(2\pi)^2}
 \int_{0}^{\infty}
  \frac{de_{0}}{e_{0}^2}\\
 &&\times\exp\left[-\frac{i}{2}\left(e_{0}m^2
 +\frac{{\Delta x}^2}{e_{0}}
 \right)
 \right]\int {\cal D}v\; \delta^4\left(\int v d\tau\right)
 \nonumber\\
 &&\times\exp\left\{i\int d\tau\left[-\frac{v^2}{2}
 -g({\sqrt e_{0}}v
 +{\Delta x})A\left({\sqrt e_{0}}\int_{0}
 ^{\tau} v(\tau^{\prime})\;d\tau^{\prime}+x_{in}
 +\tau{\Delta x}\right)\right]
 \right\}\nonumber\\
 &&\times\left[\frac{{\rm Det }\Lambda(g)}{{\rm Det }
 \Lambda(0)}
 \right]^{\frac{1}{2}}
 \left[\frac{v_{\mu}}
 {\sqrt{e_{0}}}\left(\epsilon\frac{\delta}{\delta\rho_{\mu}}
 +\frac{\partial}{\partial
 \zeta_{\mu}}\right)
 -m\left(\epsilon\frac{\delta}{\delta\rho_{5}}
 +i\gamma^5\right)\right]
 \nonumber\\
 &&\left.\left.\times\exp\left\{-\frac{1}{4}ge_{0}
 F_{\mu\nu}\frac{\partial}
 {\partial\zeta_{\mu}}\frac{\partial}
 {\partial\zeta_{\nu}}\right\}\exp\left\{\frac{1}{2}
 I \Lambda^{-1}I\right\}
 \exp(i\zeta_{\lambda}
 \gamma^{\lambda})\right.\right|_{\rho=0,\zeta=0}
 \nonumber
 \end{eqnarray}
 
 \noindent and performing functional differentiations
  with respect to 
 $\rho_{\mu}$ we get
 
 \begin{eqnarray}
 &&{\tilde S}^c(x_{out},x_{in})=-\frac{1}{2i(2\pi)^2}
 \int_{0}^{\infty}
 \frac{de_{0}}{e_{0}^2}
 \exp\left[-\frac{i}{2}\left(e_{0}m^2+\frac{{\Delta x}^2}
 {e_{0}}\right)
 \right]\int {\cal D}v\;\delta^4
 \left(\int v d\tau\right)\nonumber\\
 &&\times\exp\left\{i\int d\tau \left[-\frac{v^2}{2}
 -g({\sqrt e_{0}}v
 +{\Delta x})A\left({\sqrt e_{0}}
 \int_{0}^{\tau} v(\tau^{\prime})d\tau^{\prime}+x_{in}
 +\tau{\Delta x}\right)
 \right]\right\}\nonumber\\
 &&\times\left[\frac {{\rm Det} \Lambda(g)}{{\rm Det}
  \Lambda(0)}
 \right]^{\frac{1}{2}}\\
 &&\left.\times\left[\frac{v^{\mu}}{\sqrt{e_{0}}}K_{\mu\nu}
 \frac{\partial}{\partial
 \zeta_{\nu}}
 -im\gamma^5\right]
 \exp\left\{-\frac{1}{4}ge_{0}(FK)_{\mu\nu}
 \frac{\partial}{\partial
 \zeta_{\mu}}\frac{\partial}{\partial
 \zeta_{\nu}}\right\}\exp(i\zeta_{\lambda}\gamma^{\lambda})
 \right|_{\zeta=0}\;,\nonumber
 \end{eqnarray}
 
 \noindent where
 
 \begin{eqnarray}
 K_{\mu\nu}&=&\eta_{\mu\nu}+ge_{0}(GF)_{\mu\nu}\;,
 \nonumber\\
 G&=&\frac{1}{2}\epsilon \Lambda^{-1}\epsilon.
 \end{eqnarray}

 \noindent On the  other hand, we have

 \begin{eqnarray}
 \frac{d}{dg}{\rm Det}\Lambda(g)&=&{\rm Det}\Lambda(g)
 {\rm Tr}
 \Lambda^{-1}(g)
 \frac{d}{dg}\Lambda(g)\;,
 \nonumber\\
 \left[\frac{{\rm Det}\Lambda(g)}{{\rm Det}\Lambda(0)}
 \right]^{\frac{1}{2}}
 &=&\exp\left\{-\frac{e_{0}}{2}\int_{0}^{g}{\rm Tr}(GF)\;
  dg^{\prime}\right\}\nonumber
 \end{eqnarray}
 
 \noindent and the differentiation over the anticomutative
  variable $\zeta$
  give us a finite number of terms, so we obtain the factor
 
 \begin{eqnarray} 
 {\tilde \Phi}[v,e_{0}]&=&
 \left[\frac{v^{\mu}}{\sqrt{e_{0}}}K_{\mu\nu}\frac{\partial}
 {\partial\zeta_{\nu}}
 -im\gamma^5\right]\left[1
 -\frac{1}{4}ge_{0}(FK)_{\mu\nu}\frac{\partial}
 {\partial\zeta_{\mu}}\frac
 {\partial}{\partial\zeta_{\nu}}
 \right.\\
 &&\left.-\frac{1}{16}g^2e_{0}^2(FK)_{\mu\nu}(FK)^{*\mu\nu}
 \frac{\partial}
 {\partial\zeta_{0}}\frac
 {\partial} {\partial\zeta_{1}}\frac{\partial}{\partial
 \zeta_{2}}
 \frac{\partial}{\partial\zeta_{3}}\right]
 \nonumber\\
 &&\times\left[1+i\zeta_{\alpha}\gamma^{\alpha}
 -\frac{i}{2!}
 \zeta_{\alpha}\zeta_{\beta}\gamma^{\alpha}
 \gamma^{\beta}+\frac{i}{3!}\zeta_{\alpha}\zeta_{\beta}
 \zeta_{\sigma}
 \gamma^{\alpha}\gamma^{\beta}
 \gamma^{\sigma}+\zeta_{0}\zeta_{1}\zeta_{2}\zeta_{3}
 \gamma^5\right]
 \nonumber\\
 &&\left.\times\exp\left\{-\frac{e_{0}}{2}\int_{0}^{g}
 dg^{\prime}\;
 {\rm Tr}(GF)\right\}\right|_{\zeta=0}\;,
 \nonumber
 \end{eqnarray}
 
 \noindent with
 
 \[
 (FK)^{*\mu\nu}=\frac{1}{2}\epsilon^{\mu\nu\alpha\beta}
 (FK)_{\alpha\beta}.
 \]
 
 \noindent  In this way we have obtained the expression
  for the Dirac 
 propagator
  in an external field as a path integral over velocities,
 
 \begin{eqnarray}
 S(x_{out},x_{in})&=&\frac{1}{2(2\pi)^2}\int_{0}^{\infty}
 \frac{de_{0}}
 {e_{0}^2}
 \exp\left[-\frac{i}{2}\left(e_{0}m^2+
 \frac{{\Delta x}^2}{e_{0}}\right)\right]\\
 &&\times\int {\cal D}v\; \Phi[v,e_{0}]\; \delta^4
 \left(\int v 
 d\tau\right)\nonumber\\
 &&\times\exp\left\{i\int\left[-\frac{v^2}{2}
 -g({\sqrt e_{0}}v
 +{\Delta x})A
 \left({\sqrt e_{0}}\int_{0}^{\tau} v(\tau^{\prime})
 d\tau^{\prime}+x_{in}
 +\tau{\Delta x}\right)\right]d\tau
 \right\}\;,\nonumber
 \end{eqnarray}
 
 \noindent where
 
 \begin{eqnarray} 
&&\Phi[v,e_{0}]=\\
&&\left\{m+\frac{1}{2{\sqrt e_{0}}}v K(2-g e_{0}FK)
\gamma-\frac{i}{4}
mge_{0}(FK)_{\mu\nu}\sigma^{\mu\nu}\right.\nonumber\\
&&\left.-\frac{i}{4}g{\sqrt e_{0}}v K\gamma(FK)_{\mu\nu}
\sigma^{\mu\nu}
+\frac{1}{16}mg^2e_{0}^2
(FK)_{\mu\nu}^{*}(FK)^{\mu\nu}\gamma^5\right\}\nonumber\\
&&\times\exp\left\{-\frac{e_{0}}{2}\int_{0}^{g}dg^{\prime}\;
{\rm Tr}(GF)\right\}\;,\nonumber
\end{eqnarray}

\noindent can be understood as a spin factor in the
 representation over
 the velocities.

Now, for the particle in a constant electromagnetic field 
$ A_{\mu}=
-\frac{1}{2}F_{\mu\nu}x^{\nu}$, we can 
calculate the propagation function exactly. After the 
shift $ v 
\rightarrow v-\frac{{\Delta x}}
{{\sqrt e_{0}}}$, we use an integral representation
 for the delta function 
and obtain

\begin{eqnarray}
&&S^c(x_{out},x_{in})=\frac{1}{8\pi^2}\int_{0}^{\infty}
\frac{de_{0}}{e_{0}^2}
\exp\left[-\frac{i}{2}e_{0}m^2\right]\\
&&\times\int {\cal D}v\; \Phi[v,e_{0}]\int d\sigma\exp
\left\{-i\frac{\sigma.{\Delta x}}{{\sqrt e_{0}}}\right\}
\nonumber\\
&&\times\exp\left\{-\frac{i}{2}\int v^{\mu}(\tau)L_{\mu\nu}
(\tau,\tau^{\prime})v^{\nu}(\tau^{\prime})d\tau
\; d\tau^{\prime}\right\}\nonumber\\
&&\times\exp\left\{i\int \left[\frac{g{\sqrt e_{0}}}{2}
(Fx_{in})_{\mu}
+p_{\mu}\right]v^{\mu}d\tau\right\}\;,\nonumber
\end{eqnarray}

\noindent where

\begin{equation}
L_{\mu\nu}(\tau,\tau^{\prime})=\eta_{\mu\nu}
\delta(\tau-\tau^{\prime})
-\frac{1}
{2}ge_{0}F_{\mu\nu}\epsilon(\tau-\tau^{\prime}).
\end{equation}

\noindent Introducing the sources $j_{\mu}$ related
 to $v_{\mu}$ and 
defining $\pi=p+\frac{1}{2}
g{\sqrt e_{0}}(Fx_{in})$ plus the usual shift on
 $v$, we get

\begin{eqnarray}
&&S^c(x_{out},x_{in})=\frac{1}{8\pi^2}\int_{0}^{\infty}
\frac{de_{0}}
{e_{0}^2}
\exp\left[-\frac{i}{2}e_{0}m^2\right]\\
&&\times\int {\cal D}v\;\Phi\left[\frac{\delta}
{i\delta j},e_{0}\right]
\exp\left\{-\frac{i}{2}
\int v^{\mu}L_{\mu\nu}v^{\nu}d\tau\;
 d\tau^{\prime}\right\}\nonumber\\
&&\times\int d\pi \exp\left\{\frac{i}{2}\pi_{\mu}
Q^{\mu\nu}\pi_{\nu}
+i\pi_{\mu}\
f^{\mu}\right\}\nonumber\\
&&\times\left.\exp\left\{\frac{i}{2}gx_{out}Fx_{in}
\right\}\right|_{j=0}\;,
\nonumber
\end{eqnarray}
 
 \noindent where our compact notation means
 
 \[
 \pi_{\mu}Q^{\mu\nu}\pi_{\nu}=\int \pi_{\mu}(\tau)
 (L^{-1})^{\mu\nu}
 (\tau,\tau^{\prime})\pi_{\nu}(\tau^{\prime})
 d\tau\;d\tau^{\prime}
 \]
 
 \noindent and
 
 \[
 \pi_{\mu}f^{\mu}=\int \pi_{\mu}(\tau)(L^{-1})^{\mu\nu}
 (\tau,\tau^{\prime})j_{\nu}(\tau^{\prime})d\tau\;
  d\tau^{\prime}.
 \]
 
 \noindent After shifting the $\pi$ variables we get 
 
 \begin{eqnarray}
 &&S^{c}(x_{out},x_{in})=\frac{1}{8\pi^2}\int_{0}^{\infty}
  \frac{de_{0}}
 {e_{0}^2}
 \exp\left[-\frac{i}{2}e_{0}m^2\right]
 \int {\cal D}v\; \Phi\left[\frac{\delta}{i\delta j},e_{0}
 \right]\\
 &&\times\exp\left\{-\frac{i}{2}\int v^{\mu}L_{\mu\nu}
 v^{\nu}d\tau\; d\tau^{\prime}\right\}
 \int d\pi \exp\left\{ \frac{i}{2}\pi Q \pi \right\}
 \nonumber\\
 &&\times\exp\left\{-\frac{i}{2}\left(\frac{{\Delta x}}
 {{\sqrt e_{0}}}
 -\int_{0}^1 L^{-1}
 (\tau,\tau^{\prime})j(\tau^{\prime})d\tau\; 
 d\tau^{\prime}\right)\right.
 \nonumber\\
 &&\left.\left.\times Q^{-1}\left(\frac{{\Delta x}}
 {{\sqrt e_{0}}}
 -\int_{0}^1
  L^{-1}(\tau,\tau^{\prime})
 j(\tau^{\prime})d\tau\; d\tau^{\prime}\right)
 \right\}\right|_{j=0}.
 \nonumber
 \end{eqnarray}
 
 \noindent On the other hand, we have that ( see appendix )
 
 \[
 Q=\int Q(\tau^{\prime})d\tau^{\prime}=\frac{2}
 {ge_{0}F}\tanh
 \frac{ge_{0}F}{2}
 \]
 
 \noindent and
 
 \[
 \int L^{-1}(\tau,\tau^{\prime})j(\tau^{\prime})d\tau
 \; d\tau^{\prime}= 
 \frac{e^{\frac{ge_{0}F}{2}}}
 {\cosh\frac{ge_{0}F}{2}}\int_{0}^1 e^{-ge_{0}F\tau^{\prime}}
 j(\tau^{\prime})d\tau^{\prime}.
 \]

 \noindent So, performing the integration over $ v$ and $\pi$
 
 \begin{eqnarray}
&&S^c(x_{out},x_{in})=\frac{1}{8\pi^2}\int_{0}^{\infty}
\frac{de_{0}}{e_{0}^2}
  \; \Phi\left[\frac{\delta}{i\delta j},e_{0}\right]\\
 &&\times\left[\frac{{\rm det}\; L(g)}{{\rm det}\; L(0)}
 \right]^{-\frac{1}{2}}
 \left[\frac{{\rm det}\; Q(g)}{{\rm det }\; Q(0)}
 \right]^{-\frac{1}{2}}
 \nonumber\\
 &&\times\exp\left\{\frac{i}{2}gx_{out}Fx_{in}
 -\frac{i}{2}e_{0}m^2-\frac{i}{2e_{0}}{\Delta x}
 Q^{-1}{\Delta x}\right\}
 \nonumber\\
 &&\left.\times\exp\left\{\frac{i}{2}\int_{0}^1 j(\tau)
 K(\tau,\tau^{\prime})
 j(\tau^{\prime})d\tau\; 
 d\tau^{\prime}+i\int_{0}^1 a(\tau)j(\tau)
 d\tau\right\}
 \right|_{j=0}
 \;,\nonumber
 \end{eqnarray}
 
 \noindent where 
 
 \begin{eqnarray}
 &&K(\tau,\tau^{\prime})=\delta(\tau-\tau^{\prime})
 +\frac{1}{2}
 ge_{0}F e^{ge_{0}F(\tau-\tau^{\prime})}
 \left[\epsilon(\tau-\tau^{\prime})-\coth
 \frac{ge_{0}F}{2}
 \right]\;,\\
 &&a(\tau)=\frac{ge_{0}{\Delta x}}{2{\sqrt e_{0}}}
 \left(1+\coth\frac{ge_{0}F}{2}\right)e^{-ge_{0}F\tau}.
 \nonumber
 \end{eqnarray}

 \noindent Finally, the propagator can be
  written as 
 
 \begin{eqnarray}
 &&S^c(x_{out},x_{in})=\frac{1}{32\pi^2}
 \int_{0}^{\infty}
 de_{0}
 \left[{\rm det}\frac
 {\sinh\frac{ge_{0}F}{2}}{gF}\right]^{-\frac{1}{2}}
 \Phi[a,e_{0}]\\
 &&\times\exp\left\{\frac{i}{2}gx_{out}Fx_{in}
 -\frac{i}{2}e_{0}
 m^2-\frac{i}{4}(x_{out}-x_{in})gF\coth
\left(\frac{ge_{0}F}{2}\right)(x_{out}-x_{in})
\right\}\;,
\nonumber
\end{eqnarray}

\noindent where

\begin{eqnarray}
&&\Phi[a,e_{0}]=
\left[m+\frac{1}{2{\sqrt e_{0}}}aK(2-ge_{0}FK)\gamma
-\frac{i}{4}mge_{0}
(FK)_{\mu\nu}\sigma^{\mu\nu}\right.\\
&&\left.-\frac{i}{4}g{\sqrt e_{0}}aK\gamma(FK)_{\mu\nu}
\sigma^{\mu\nu}
+\frac{1}{16}mg^2e_{0}^2(FK)_{\mu\nu}^{*}(FK)^{\mu\nu}
\gamma^5\right]
\nonumber\\
&&\times\exp\left\{-\frac{e_{0}}{2}
\int_{0}^g{\rm Tr}(FG) 
dg^{\prime}\right\}\;,\nonumber
\end{eqnarray}

\noindent is the spin factor which does not depend
 on the trajectories.
 Having in mind the following notation

\begin{eqnarray}
&&\sigma^{\mu\nu}=\frac{i}{2}[\gamma^{\mu},
\gamma^{\nu}]\;,\;
\gamma^5=\gamma^0\gamma^1\gamma^2\gamma^3
\;,\nonumber\\
&& B_{\mu\nu}=F_{\mu\lambda}*K^{\lambda}_{\nu}\;,\;
B^{*\mu\nu}=\frac{1}{2}\epsilon^{\mu\nu\alpha\beta}
B_{\alpha\beta}\;,\nonumber\\
&& K_{\mu\nu}=\eta_{\mu\nu}+ge_{0}G_{\mu\lambda}
(g)*F^{\lambda}_{\nu}\;,\;
G_{\mu\nu}(g)=\frac{1}{2}
\epsilon *\Lambda^{-1}_{\mu\nu}(g) * \epsilon\;,\\
&& \Lambda^{-1}_{\mu\lambda}(g)*\Lambda^{\lambda\nu}(g)
=\delta_{\mu}^{\nu}
\delta(\tau-\tau^{\prime})\;,\;
{\cal F}_{\mu\nu}(\tau,\tau^{\prime})=
F_{\mu\nu}(x(\tau))\delta(\tau-\tau^{\prime})
\;,\nonumber\\
&& \Lambda_{\mu\nu}(g)=\eta_{\mu\nu}\epsilon
-\frac{1}{2}
ge_{0}\epsilon * {\cal F}_{\mu\nu} * \epsilon
\;,\nonumber\\
&&\epsilon * {\cal F}_{\mu\nu} * \epsilon 
= \int_{0}^1 d\tau_{1}\int_{0}^1
 d\tau_{2}\;\epsilon(\tau,\tau_{1})\;
 {\cal F}_{\mu\nu}(\tau_{1},\tau_{2})\;
\epsilon(\tau_{2},\tau^{\prime})\;,\nonumber
\end{eqnarray}

\noindent where $\epsilon^{0123}=1$ and $ 
\epsilon(\tau,\tau^{\prime})$$\;$,
${\cal F}_{\mu\nu}$$\;$,$\Lambda_{\mu\nu}(g)$$\;$,  
$G_{\mu\nu}(g)$ are matrices with continous
 indices $\tau,\tau^{\prime}$, 
we rewrite the spin factor as

\begin{eqnarray}
&&\Phi[a,e_{0}]=\left[m+\frac{1}{2{\sqrt e_{0}}}
a^{\mu}*K_{\mu\lambda}
\left(2\eta^{\lambda\kappa}
-ge_{0}B^{\lambda\kappa}\right)\gamma_{\kappa}\right.\\
&&-\frac{i}{4}g\left.\left(me_{0}+a^{\mu}*K_{\mu\lambda}
\gamma^{\lambda}\right)
B_{\kappa\nu}\sigma^{\kappa\nu}
+\frac{1}{16}mg^2e_{0}^2B_{\alpha\beta}^{*}B^{\alpha\beta}
\gamma^5\right]\nonumber\\
&&\times\exp\left\{-\frac{e_{0}}{2}\int_{0}^g dg^{\prime}\;
{\rm Tr}\; G(g^{\prime})*{\cal F}\right\}.\nonumber
\end{eqnarray}

\noindent We observe that in the case of a constant 
field $G$$\;$,
$K$ and $B$ do not depend on the trajectory, so 
 
 \begin{eqnarray}
 &&G=\frac{1}{2}\left[I\epsilon(\tau-\tau^{\prime})
 -\tanh\frac{ge_{0}F}{2}
 \right]\exp\left\{ge_{0}F
 (\tau-\tau^{\prime})\right\}\;,\nonumber\\
 &&K=\left(I-\tanh\frac{ge_{0}F}{2}\right)
 \exp(ge_{0}F\tau)\;,\\
 &&B=\frac{2}{ge_{0}}\tanh\frac{ge_{0}F}{2}
 \;,\nonumber
 \end{eqnarray}
 
 \noindent where $I$ is the unit $4 \times 4$
  matrix. Using this latter
  expression, the function
 $a(\tau)$ defined earlier and integrating over
  $\tau$, we can show that 
 the spin factor is given by
 
 \begin{eqnarray}
 &&\Phi[x_{out},x_{in},e_{0}]=
 \left[m+\frac{g}{2}(x_{out}-x_{in})F\left(\coth
 \frac{ge_{0}F}{2}-1\right)
 \gamma\right]\\
 &&\times\left[{\rm det}\cosh
 \frac{ge_{0}F}{2}\right]^{\frac{1}{2}}
 \left\{1-\frac{i}{2}\left(\tanh\frac
 {ge_{0}F}{2}\right)_{\mu\nu}\sigma^{\mu\nu}
 \right.\nonumber\\
 &&+\left.\frac{1}{8}\epsilon^{\alpha\beta\mu\nu}
 \left(\tanh\frac{ge_{0}F}{2}
 \right)_{\alpha\beta}\left(\tanh\frac{ge_{0}F}{2}
 \right)_{\mu\nu}\gamma^5\right\}
 .\nonumber
 \end{eqnarray}
 
 \noindent This result has been associated with
  a factor that came from 
 the Schwinger formula for the spinning particle
  propagator
 in a constant electromagnetic field and these 
 look equivalent\cite{R4}. 
 The propagator is

 \begin{eqnarray}
 &&S^c(x_{out},x_{in})=\frac{1}{32\pi^2}
 \int_{0}^{\infty}de_{0}
 \left[{\rm det}\frac{\sinh\frac{ge_{0}F}{2}}
 {gF}\right]^{-\frac{1}{2}}\Phi[x_{out},x_{in},e_{0}]\\
 &&\times\exp\left\{\frac{i}{2}gx_{out}Fx_{in}
 -\frac{i}{2}
 e_{0}m^2-\frac{i}{4}(x_{out}-x_{in})gF
 \coth\frac{ge_{0}F}{2}(x_{out}-x_{in})\right\}
 .\nonumber
 \end{eqnarray}
 
 \noindent The causal Green function of the 
 Dirac-Pauli equation is the 
 propagator of the relativistic spinning particle
  with anomalous magnetic 
 moment\cite{R6}:
 
 \begin{eqnarray}
 &&{\tilde S^c}(x_{out},x_{in})=\exp\left(i\gamma^n
 \frac{\partial_{l}}
 {\partial\theta^n}\right)\int_{0}^{\infty} de_{0}
 \int d\chi_{0}\\
 &&\times\int\exp\left(i\left\{\int_{0}^1
 \left[-\frac{{\dot x}^2}{2e}-\frac
 {e{\cal M}^2}{2}-{\dot x}^{\alpha}
 \left(gA_{\alpha}(x)+4i\mu\psi^5
 F_{\alpha\beta}(x)\psi^{\beta}\right)
 \right.\right.\right.\nonumber\\
 &&\left.+igeF_{\alpha\beta}(x)\psi^{\alpha}\psi^{\beta}
 +i\left(\frac
 {{\dot x}_{\alpha}\psi^{\alpha}}{e}
 -{\cal M}^{*}\psi^5\right)\chi
 -i\psi_{n}{\dot \psi}^n+\pi{\dot e}
 +\nu{\dot \chi}
 \right]d\tau\nonumber\\
 &&\left.\left.\left.+
 \psi_{n}(1)\psi^n(0)\right\}\right)M(e)\; Dx\; De\;
  D\pi\; D\chi\;
  D\nu\;{\cal D}\psi\right|_{\theta=0}\;,\nonumber
 \end{eqnarray}
 
 \noindent where $ {\cal M}^{*}=m+2i\mu F_{\alpha\beta}
 \psi^{\alpha}\psi^{\beta}$
  and $M(e)$ is defined as in the former case.
  
  Performing the integration over $\pi$ and $\nu$ we obtain
  
  \begin{eqnarray}
  &&{\tilde S}^c(x_{out},x_{in})=\exp\left(i\gamma^n
  \frac{\partial_{l}}
  {\partial\theta^n}\right)\int_{0}^{\infty}de_{0}\;
   Dp\; d\chi_{0}\;Dx\;
   {\cal D}\psi\\
   &&\times\exp\left\{i\int_{0}^1\left[-\frac{{\dot x}^2}
   {2e_{0}}-\frac{e_{0}
   m^2}{2}-g{\dot x}A+\frac{e_{0}p^2}{2}\right]d\tau
   \right\}\nonumber\\
   &&\times\exp\left\{i\int_{0}^1\left[2e_{0}\mu^{2}
   F_{\alpha\beta}
   \psi^{\alpha}\psi^{\beta}F_{\sigma\rho}\psi^{\sigma}
   \psi^{\rho}
   +2ie_{0}\mu mF_{\alpha\beta}\psi^{\alpha}
   \psi^{\beta}\right.\right.\nonumber\\
   &&-4i\mu{\dot x}^{\alpha}\psi^5 F_{\alpha\beta}
   \psi^{\beta}+ige_{0}
   F_{\alpha\beta}\psi^{\alpha}\psi^{\beta}\nonumber\\
   &&\left.\left.\left.+i\left(\frac{{\dot x}_{\alpha}
   \psi^{\alpha}}
   {e_{0}}-{\cal M}^*\psi^5
   \right)\chi_{0}-i\psi_{n}{\dot \psi}^n\right]d\tau
   +\psi_{n}(1)\psi^n(0)
   \right\}\right|_{\theta=0}.\nonumber
   \end{eqnarray}
   
   \noindent After changing the variables $x$ by $v$ 
   in the same way as before, 
   we associate the sources $j$ to $v$ and get
   
   \begin{eqnarray}
   &&{\tilde S}^c(x_{out},x_{in})=\frac{1}{i(2\pi)^2}
   \exp\left(i\gamma^n
   \frac{\partial_{l}}{\partial\theta^n}\right)
   \int_{0}^{\infty}
    de_{0}\; I(e_{0})\\
   &&\times\exp\left\{i\int_{0}^1d\tau\left[2e_{0}
   \mu^2F_{\alpha\beta}
   \psi^{\alpha}\psi^{\beta}F_{\sigma\rho}\psi^{\sigma}
   \psi^{\rho}
   \right.\right.\nonumber\\
   &&+2ie_{0}\mu m F_{\alpha\beta}\psi^{\alpha}\psi^{\beta}
   -4i\mu{\sqrt e_{0}}
   a^{\alpha}(\tau)\psi^5F_{\alpha\beta}\psi^{\beta}
   +ige_{0}F_{\alpha\beta}
   \psi^{\alpha}\psi^{\beta}\nonumber\\
   &&\left.\left.\left.+i\left(\frac{a_{\alpha}}
   {{\sqrt e_{0}}}\psi^{\alpha}
   -{\cal M}^*\psi^5
   \right)\chi_{0}-i\psi_{n}{\dot \psi}^n\right]
   +\psi_{n}(1)\psi^{n}(0)
   \right\}
   d\chi_{0}\; {\cal D}\psi\right|_{\theta=0}\;,\nonumber
   \end{eqnarray}
   
   \noindent where $I(e_{0})$ is obviously defined.
   
   In this stage, we substitute the $\psi$ by $\omega$
    variables 
   and the result is
   
   \begin{eqnarray}
   &&{\tilde S}^c(x_{out},x_{in})=\frac{1}{i(2\pi)^2}
   \exp\left(i\gamma^n\frac
   {\partial_{l}}{\partial\theta^n}\right)
   \int_{0}^{\infty}de_{0}\; 
   I(e_{0})\\
   &&\times\exp\left\{i\int\right.\left[2e_{0}\mu^2
   \left(F_{\alpha\beta}
   \psi^{\alpha}
   \psi^{\beta}\right)^2+2ie_{0}\mu m F_{\alpha\beta}
   \psi^{\alpha}\psi^{\beta}
   \right.\nonumber\\
   &&\left.\left.+i\left(\frac{a_{\alpha}}{{\sqrt e_{0}}}
   \psi^{\alpha}-{\cal M}^{*}\psi^5
   \right)\chi_{0}\right]d\tau\right\}\nonumber\\
   &&\times\int {\cal D}\omega \exp\left\{-\frac{1}{2}\int
    \omega^{n}(\tau)
   \Lambda_{nm}(\tau,\tau^{\prime})\omega^m(\tau^{\prime})
    d\tau\;d\tau^{\prime}
   \right.\nonumber\\
   &&\left.\left.+\int \zeta_{n}\omega^n d\tau
   -\mu{\Delta x}^{\alpha}\theta^5
   F_{\alpha\beta}\theta^{\beta}-\frac{1}{4}ge_{0}
   F_{\alpha\beta}
   \theta^{\alpha}\theta^{\beta}\right\}
   \right|_{\theta=0}\;,\nonumber
   \end{eqnarray}
   
   \noindent with
   
   \begin{eqnarray}
   &&\Lambda_{\alpha\beta}=\epsilon(\tau-\tau^{\prime})
   \eta_{\alpha\beta}
   -\frac{1}{2}ge_{0}\epsilon F_{\alpha\beta} \epsilon
   \;,\nonumber\\
   &&\Lambda_{55}=\epsilon(\tau-\tau^{\prime})\eta_{55}
   \;,\; \eta_{55}=-1\;,
   \nonumber\\
   &&{\cal D}\omega=\frac{D\omega}{\int D\omega\; 
   e^{-\frac{1}{2}
   \omega^n
   \epsilon \omega_{n} }}\;,\nonumber\\
   &&\zeta_{n}={\tilde Q}_{n}+c_{n}\;,\\
   &&{\tilde Q}_{5}=0,\;\;{\tilde Q}_{\alpha}=\frac{1}{2}
   ge_{0}\int
   \epsilon(\tau^{\prime\prime}-\tau)F_{\alpha\beta}
   \theta^{\beta}d\tau\;,
   \nonumber\\
   &&c_{5}=-\mu{\sqrt e_{0}}\int a^{\alpha}
   (\tau^{\prime\prime})
   F_{\alpha\beta}\;\epsilon(\tau^{\prime\prime}-\tau)\;
   \theta^{\beta} 
   d\tau^{\prime\prime}\;,\nonumber\\
   &&c_{\beta}=\mu{\sqrt e_{0}}\int a^{\alpha}
   (\tau^{\prime\prime})
   F_{\alpha\beta}\;\epsilon(\tau^{\prime\prime}
   -\tau)\theta^5
   d\tau^{\prime\prime}.\nonumber
   \end{eqnarray}
   
   \noindent Therefore,
   
   \begin{eqnarray}
   \label{e2}
   &&{\tilde S}^c(x_{out},x_{in})=\frac{1}{i(2\pi)^2}
   \exp\left(i\gamma^n
   \frac{\partial_{l}}{\partial\theta^n}\right)
   \int_{0}^{\infty}de_{0}\;
   I(e_{0})\int {\cal D}\omega\;d\chi_{0}\\
   &&\times\exp\left\{i\left(\frac{a_{\alpha}}
   {{\sqrt e_{0}}}
   \psi^{\alpha}
   -{\cal M}^*\psi^5\right)\chi_{0}\right\}
   \nonumber\\ 
   &&\times\exp\left\{-\frac{1}{2}\omega^n
   \left(\Lambda_{nm}
   +4e_{0}\mu m 
   {\cal L}_{nm}\right)\omega^m+\left(\zeta_{n}-2e_{0}
   \mu m Q_{n}\right)
   \omega^n\right.\nonumber\\
   &&\left.-\frac{1}{4}e_{0}(g+2m\mu)F\theta\theta
   -\mu{\Delta x}^{\alpha}
   \theta^5 F_{\alpha\beta}\theta^{\beta}\right\}
   \nonumber\\
   &&\times\exp\left\{2ie_{0}\mu^2
   \int{\cal L}_{nm}{\cal L}_{st}\omega^n
   \omega^m\omega^s\omega^t d\tau\;d\tau^{\prime}
   \;d\tau^{\prime\prime}\;
   d\tau^{\prime\prime\prime}\right.\nonumber\\
   &&+4ie_{0}\mu^2\int{\cal L}_{nm}Q_{s}\omega^n
   \omega^m\omega^sd\tau\;
   d\tau^{\prime}\;d\tau^{\prime\prime}\nonumber\\
   &&-2ie_{0}\mu^2\int Q_{nm}\omega^n\omega^md\tau
   \;d\tau^{\prime}+ie_{0}\mu^2
   F\theta\theta\int\omega^n{\cal L}_{nm}\omega^m
    d\tau\;d\tau^{\prime}
   \nonumber\\
   &&\left.\left.+ie_{0}\mu^2 F\theta\theta\int Q_{n}\omega^n
    d\tau+\frac{i}{8}e_{0}\mu^2
   (F\theta\theta)^2\right\}\right|_{\theta=0}\;,\nonumber
   \end{eqnarray}
   
   \noindent where
   
   \begin{eqnarray}
   &&{\cal L}_{\alpha\beta}=-\frac{1}{4}\int\epsilon(\tau-
   \tau^{\prime\prime})F_{\alpha\beta}\epsilon
   (\tau^{\prime\prime}-
   \tau^{\prime})d\tau^{\prime\prime}\;,\nonumber\\
   &&{\cal L}_{55}=0={\cal L}_{5\beta}\;,\\
   &&Q_{\alpha}=-\frac{1}{2}\int\epsilon
   (\tau^{\prime\prime}-\tau)
   F_{\alpha\beta}\theta^{\beta}
   d\tau^{\prime\prime}\;,\;Q_{5}=0.\nonumber
   \end{eqnarray}
   
   \noindent Consider
   
   \[
   \exp\left\{-\frac{1}{2}\omega^n{\tilde 
   \Lambda}_{nm}\omega^m+I_{n}
   \omega^n\right\}
   \]
   
   \noindent with
   
   \[
   {\tilde \Lambda}_{nm}=\Lambda_{nm}
   +4e_{0}m\mu{\cal L}_{nm}
   \]
   
   and
   
   \[
   I_{n}=\rho_{n}+\zeta_{n}-2e_{0}m\mu Q_{n}
   +ie_{0}\mu^2 F\theta\theta Q_{n}
   \;,
   \]
   
   \noindent where the sources $\rho$ related
    to $\omega$ velocities were 
   inserted and after shifting it we get
   
   \[
   \exp\left\{-\frac{1}{2}\omega{\tilde \Lambda}
   \omega+\frac{1}{2}I
   {\tilde \Lambda}^{-1}I\right\}.
   \]
   
   \noindent Going back to expression(\ref{e2}) 
   and performing the
    integration over 
   $\chi_{0}$, taking into account the identity
    defined earlier for 
   Grassmann algebra, we obtain
   
   \begin{eqnarray}
   &&{\tilde S}^c(x_{out},x_{in})=-\frac{1}{2i(2\pi)^2}
   \int{\cal D}\omega
   \;de_{0}\;I(e_{0})\left[\frac{a^{\alpha}}{{\sqrt e_{0}}}
   \left(\epsilon
   \frac{\delta}{\delta\rho_{\alpha}}+\frac{\partial}
   {\partial\zeta_{\alpha}}
   \right)\right.\\
   &&-m\left.\left(\epsilon\frac{\delta}{\delta\rho_{5}}
   +i\gamma^5\right)-i\mu
    F_{\alpha\beta}\left(\epsilon\frac{\delta}
    {\delta\rho_{\alpha}}+
    \frac{\partial}{\partial\zeta_{\alpha}}\right)
    \left(\epsilon
    \frac{\delta}{\delta\rho_{\beta}}+\frac{\partial}
    {\partial\zeta_{\beta}}
    \right)\left(\epsilon\frac{\delta}{\delta\rho_{5}}
    +i\gamma^5\right)\right]
    \nonumber\\
    &&\times\exp\left\{-\frac{1}{2}\omega{\tilde \Lambda}
    \omega\right\}\exp
    \left\{-i\mu{\Delta x}^{\alpha}\gamma^5 F_{\alpha\beta}
    \frac{\partial}
    {\partial\zeta_{\beta}}-\frac{e_{0}}{4}(g+2m\mu)
    F_{\mu\nu}\frac
    {\partial}{\partial\zeta_{\mu}}\frac{\partial}
    {\partial\zeta_{\nu}}
    \right\}\nonumber\\
    &&\times\exp\left\{-2ie_{0}\mu^2 Q_{\mu}
    Q_{\nu}\frac{\partial}
    {\partial\rho_{\mu}}\frac{\partial}{\partial
    \rho_{\nu}}\right\}
    \exp\left\{ie_{0}\mu^2 F_{\alpha\beta}\frac{\partial}
    {\partial
    \zeta_{\alpha}}\frac{\partial}{\partial\zeta_{\beta}}
    {\cal L}_{\mu\nu}
    \frac
    {\partial}{\partial\rho_{\mu}}\frac{\partial}{\partial
    \rho_{\nu}}
    \right\}\nonumber\\
    &&\times\exp\left\{\frac{i}{8}e_{0}\mu^2\left(F_{\mu\nu}
    \frac{\partial}
    {\partial\zeta_{\mu}}\frac{\partial}{\partial\zeta_{\nu}}
    \right)^2
    \right\}\nonumber\\
    &&\times\left.\exp\left\{\frac{1}{2}I{\tilde 
    \Lambda}^{-1}I\right\}\exp
    \left(i\zeta_{\lambda}\gamma^{\lambda}\right)
    \right|_{\zeta=0,\rho=0}.
    \nonumber
    \end{eqnarray}
    
    \noindent We underline that all the following 
    steps are the same as in the case without anomalous 
    magnetic moment but the expressions are a little
     more entangled; however, we define some terms that appear
     in our calculation ( see appendix )

    \begin{eqnarray}
    &&{\tilde g}=\frac{1}{2}(g+2m\mu)\;,\nonumber\\
    &&G(\tau,\tau^{\prime})=\frac{1}{2}e^{2e_{0}
    {\tilde g}F(\tau
    -\tau^{\prime})}
    [\epsilon(\tau-\tau^{\prime})-\tanh{\tilde g}
    e_{0}F]\;,\nonumber\\
    &&K_{\mu\sigma}=\eta_{\mu\sigma}+e_{0}g(GF)_{\mu\sigma}
    \;,\nonumber\\
    &&G^{55}=\frac{1}{2}\epsilon({\tilde \Lambda}^{-1})^{55}
    \epsilon\;,\\
    &&{\tilde \Lambda}_{\alpha\beta}=\epsilon\eta_{\alpha\beta}
    -\frac{1}{2}
    (g+2m\mu)\epsilon F_{\alpha\beta}\epsilon
    \;,\nonumber\\
    &&{\tilde \Lambda}_{55}=\epsilon\eta_{55}\;\;,
    \eta_{55}=-1\;,\nonumber\\
    &&{\tilde \Lambda}_{\alpha\beta}^{-1}=\frac{1}{2}
    \frac{\partial}
    {\partial\tau}\delta(\tau-\tau^{\prime})
    +\frac{\left[(g+2m\mu)e_{0}
    F\right]^2}{4}e^{(g+2m\mu)e_{0}F(\tau-\tau^{\prime})}
    \left[\epsilon(\tau-\tau^{\prime})-\tanh
    \frac{(g+2m\mu)e_{0}F}{2}\right]
    \nonumber\\
    &&+\frac{(g+2m\mu)e_{0}F}{2}e^{(g+2m\mu)e_{0}F}
    \delta(\tau-\tau^{\prime}).\nonumber
    \end{eqnarray}
    
    \noindent The final result for the spinning
     particle propagator with 
    anomalous magnetic moment in a constant
     electromagnetic field is presented 
    up to second order in $g+2m\mu$
     and for $\mu\rightarrow 0$ we 
    see that our expression coincides with the propagator 
    of the spinning particle obtained before:

    \begin{eqnarray}
    &&S^c(x_{out},x_{in})=\frac{1}{2(2\pi)^2}
    \int_{0}^{\infty}de_{0}\left[
    {\rm det}\frac{\sinh\frac{ge_{0}F}{2}}
    {\frac{gF}{2}}\right]^{-\frac{1}{2}}
    \\
    &&\times\varphi[x_{out},x_{in},e_{0}]
    \exp\left\{\frac{i}{2}
    \left(gx_{out}Fx_{in}
    -e_{0}m^2-\frac{1}{2}{\Delta x}gF\coth
    \frac{ge_{0}F}{2}{\Delta x}\right)
    \right\}\;,\nonumber
    \end{eqnarray}

    \noindent where the spin factor is
    
    \begin{eqnarray}
    \varphi[x_{out},x_{in},e_{0}]&=&\left\{{\rm det}
    \coth\left(\frac{g+2m\mu}{2}
    \right)e_{0}F\right\}^{\frac{1}{2}}\left[m
    +\frac{1}{2{\sqrt e_{0}}}
    aK\gamma
    [2-(g+2m\mu)e_{0}(FK)]\right.\nonumber\\
    &&-\frac{i}{4}m(g+2m\mu)e_{0}(FK)_{\mu\nu}
    \sigma^{\mu\nu}
    -\frac{i}{4}(g+2m\mu)
    {\sqrt e_{0}}aK\gamma(FK)_{\mu\nu}\sigma^{\mu\nu}
    \nonumber\\
    &&+\left.\frac{1}{16}me_{0}^2(g+2m\mu)^2
    (FK)_{\mu\nu}^{*}(FK)^{\mu\nu}
    \gamma^5\right]\nonumber\\
    &=&\left\{m+\frac{1}{2}(g+2m\mu)(x_{out}
    -x_{in})F\left
    [\coth\left(\frac{g+2m\mu}{2}\right)e_{0}F
    -1\right]\gamma\right\}
    \\
    &&\times\left[{\rm det}\coth\left
    (\frac{g+2m\mu}{2}\right)e_{0}F\right]
    ^{\frac{1}{2}}\left\{1-\frac{i}{2}\left
    [\tanh\left(\frac{g+2m\mu}{2}
    \right)e_{0}F\right]_{\mu\nu}\sigma^{\mu\nu}
    \right.\nonumber\\
    &&\left.+\frac{1}{8}\epsilon^{\alpha\beta\mu\nu}
    \left[
    \tanh\left(\frac{g+2m\mu}{2}\right)
    e_{0}F\right]_{\alpha\beta}\left[\tanh\left
    (\frac{g+2m\mu}{2}\right)e_{0}F
    \right]_{\mu\nu}\gamma^5\right\}.\nonumber
    \end{eqnarray}
    
    \noindent  The propagator can be put in the
     Schwinger representation 
    \cite{R7}, taking into account the equivalence
     established in\cite{R4}:
    
    \begin{eqnarray}
    &&S^c(x_{out},x_{in})=\frac{1}{32\pi^2}
    \int_{0}^{\infty}
     de_{0}\left[{\rm det}
    \frac{\sinh\frac{ge_{0}F}{2}}{gF}
    \right]^{-\frac{1}{2}}\\
    &&\times\exp\left\{\frac{i}{2}\left(gx_{out}
    Fx_{in}-e_{0}m^2-\frac{1}{2}
    {\Delta x}gF\coth\frac{ge_{0}F}{2}{\Delta x}
    \right)\right\}\nonumber\\
    &&\times\left\{m+\frac{1}{2}(g+2m\mu){\Delta x}F
    \left[\coth\left(\frac{g+2m\mu}{2}\right)
    e_{0}F-1\right]\gamma\right\}\exp
    \left\{-\frac{i}{4}(g+2m\mu)F_{\mu\nu}
    \sigma^{\mu\nu}\right\}.\nonumber
    \end{eqnarray}

    \noindent In this paper we worked out the
     spinning particle propagator 
    with and without anomalous magnetic moment
     in a constant electromagnetic
     field. We have implemented this for the path integrals 
     over velocities and 
     we have obtained a representation involving a spin
      factor whose structure 
     is given in terms of independent $\gamma-$matrix.
      In the latter example 
     we write down by inference a Schwinger representation
      for the spinning
      particle with anomalous magnetic moment.

      {\large{\bf Acknowledgment}}
      
      I would like to thank S. Shelly Sharma for
       reading the manuscript.

      \appendix
      \section*{}

      We are going to calculate the inverse
       matrix of
      
      \begin{equation}
      L(\tau,\tau^{\prime\prime})=\delta(\tau
      -\tau^{\prime\prime})
      -\frac{\cal F}{2}
      \epsilon(\tau-\tau^{\prime\prime})\;,
      \end{equation}
      
      \noindent with ${\cal F}=ge_{0}F$.

      Given that
      
      \begin{equation}
      \int_{0}^1L_{\mu\nu}(\tau,\tau^{\prime\prime})
      (L^{-1})^{\nu\beta}
      (\tau^{\prime\prime},\tau^{\prime})d\tau^{\prime\prime}
      =\delta_{\mu}
      ^{\beta}\delta(\tau-\tau^{\prime})\;,
      \end{equation}
      
      \noindent then
      
      \begin{equation}
      L_{\mu\nu}^{-1}(\tau,\tau^{\prime})
      -\frac{{\cal F}_{\mu}^{\alpha}}{2}
      \int_{0}^1\epsilon(\tau-\tau^{\prime\prime})
      L_{\alpha\nu}^{-1}
      (\tau^{\prime\prime},\tau^{\prime})d\tau^{\prime\prime}
      =\eta_{\mu\nu}\delta(\tau-\tau^{\prime})
      \end{equation}
      
      \noindent and for $\tau=0$ we have
      
      \begin{equation}
      \label{A1}
      L_{\mu\nu}^{-1}(0,\tau^{\prime})
      +\frac{{\cal F}_{\mu}^{\alpha}}{2}
      \int_{0}^1L_{\mu\nu}^{-1}(\tau^{\prime\prime},
      \tau^{\prime})d\tau
      ^{\prime\prime}=\eta_{\mu\nu}\delta(\tau^{\prime})\;,
      \end{equation}
      
      \noindent as initial condition of the differential
       equation
      
      \begin{equation}
      \label{A2}
      \frac{\partial}{\partial\tau}L_{\mu\nu}^{-1}
      (\tau,\tau^{\prime})-
      {\cal F}_{\mu}^{\alpha}L_{\alpha\nu}^{-1}
      (\tau,\tau^{\prime})
      =\eta_{\mu\nu}\frac{\partial}{\partial\tau}
      \delta(\tau-\tau^{\prime})\;,
      \end{equation}
      
      \noindent with general solution of the type
       $L^{-1}(\tau,\tau^{\prime})
      =e^{{\cal F}\tau}
      c(\tau,\tau^{\prime})$. Substituting this
       in (\ref{A2}) we obtain 
      
      \begin{equation}
      c(\tau,\tau^{\prime})=\int_{0}^{\tau}e^{-{\cal F}
      \tau^{\prime\prime}}
      \frac{\partial}{\partial\tau^{\prime\prime}}
      \delta(\tau^{\prime\prime}
      -\tau^{\prime})d\tau^{\prime\prime}+c(\tau^{\prime})
      \end{equation}
      
      \noindent and as $L^{-1}(0,\tau^{\prime})
      =c(\tau^{\prime})$ we find
      
      \begin{eqnarray}
      \label{A3}
      L^{-1}(\tau,\tau^{\prime})&=&e^{{\cal F}\tau}
      [c(\tau^{\prime})-\delta
      (\tau^{\prime})]+\delta(\tau-\tau^{\prime})\\
      &&+{\cal F}\theta(\tau-\tau^{\prime})\exp{\cal F}
      (\tau-\tau^{\prime}).
      \nonumber
      \end{eqnarray}
      
      \noindent Substituting (\ref{A3}) into (\ref{A1})
       we get
      
      \begin{equation}
      \label{A4}
      c(\tau^{\prime})-\delta(\tau^{\prime})=-1
      \left\{1+\exp{\cal F}\right\}
      ^{-1}{\cal F}\exp{\cal F}(\tau-\tau^{\prime})\;,
      \end{equation}
      
      \noindent and with (\ref{A4}) into (\ref{A3}) we
       extract the result
      
      \begin{equation}
      L^{-1}(\tau,\tau^{\prime})=\delta(\tau-\tau^{\prime})
      +\frac{{\cal F}}{2}
      \exp{\cal F}(\tau-\tau^{\prime})\left[\epsilon
      (\tau-\tau^{\prime})
      -\tanh\frac{\cal F}{2}\right].
      \end{equation}
      
      \noindent The inverse matrix of 
      ${\tilde \Lambda}_{\alpha\beta}$
       is defined through another matrix
       
       \begin{equation}
       \label{A12}
       \Sigma_{\alpha\beta}(\tau,\tau^{\prime})
       =\int\epsilon(\tau-\lambda)
       ({\tilde \Lambda}^{-1})_{\alpha\beta}
       (\lambda,\tau^{\prime})\;d\lambda
       \end{equation}
       
       \noindent and the function $G_{\alpha\beta}$
        is given by the relation
       
       \begin{equation}
       \label{A11}
       G_{\alpha\beta}(\tau,\tau^{\prime})
       =\frac{1}{2}\int
       \Sigma_{\alpha\beta}(\tau,\lambda^{\prime})
       \epsilon(\lambda^{\prime}-
       \tau^{\prime})\;d\lambda^{\prime}
       \end{equation}

       \noindent and

       \begin{equation}
       {\tilde \Lambda}_{\alpha\beta}(\tau,\tau^{\prime})
       =\epsilon(\tau-
       \tau^{\prime})\eta_{\alpha\beta}-e_{0}{\tilde g}
       \int\epsilon(\tau-
       \tau^{\prime\prime})F_{\alpha\beta}\epsilon
       (\tau^{\prime\prime}-
       \tau^{\prime})\;d\tau^{\prime\prime}.
       \end{equation}
       
       \noindent Once we have

       \begin{equation}
       \int{\tilde \Lambda}_{\mu\nu}
       (\tau,\tau^{\prime\prime})
       ({\tilde \Lambda}^{-1})^{\nu\beta}
       (\tau^{\prime\prime},\tau^{\prime})\;
       d\tau^{\prime\prime}=\delta_{\mu}^{\beta}
       \delta(\tau-\tau^{\prime}),
       \end{equation}
       
       \noindent one has

       \begin{equation}
       \Sigma_{\mu\nu}(\tau,\tau^{\prime})
       -e_{0}{\tilde g}\int\epsilon
       (\tau-\lambda)F_{\mu}^{\alpha}
       \Sigma_{\alpha\nu}(\lambda,\tau^{\prime})
       \;d\lambda=\eta_{\mu\nu}\delta(\tau-\tau^{\prime}).
       \end{equation}
       
       \noindent This equation is equivalent
        to differential equation
       
       \begin{equation}
       \label{A5}
       \frac{\partial}{\partial\tau}
       \Sigma_{\mu\nu}(\tau,\tau^{\prime})
       -2e_{0}{\tilde g}\int\delta(\tau-\lambda)
       F_{\mu}^{\alpha}
       \Sigma_{\alpha\nu}(\lambda,\tau^{\prime})
       \;d\lambda=\eta_{\mu\nu}
       \frac{\partial}{\partial\tau}
       \delta(\tau-\tau^{\prime}),
       \end{equation}
       
       \noindent with initial condition
       
       \begin{equation}
       \label{A6}
       \Sigma_{\mu\nu}(0,\tau^{\prime})
       +e_{0}{\tilde g}\int_{0}^1 F_{\mu}
       ^{\alpha}\Sigma_{\alpha\nu}(\lambda,\tau^{\prime})
       \;d\lambda=
       \eta_{\mu\nu}\delta(\tau).
       \end{equation}
       
       \noindent By (\ref{A5}) we have
       
       \begin{equation}
       \label{A7}
       \frac{\partial}{\partial\tau}\Sigma(\tau,\tau^{\prime})
       -{\cal G}
       \Sigma(\tau,\tau^{\prime})=\frac{\partial}{\partial\tau}
       \delta(\tau-\tau^{\prime}),
       \end{equation}
       
       \noindent with ${\cal G}=2e_{0}{\tilde g}F$.
       
       Inserting the general solution $\Sigma(\tau,\tau^{\prime})
       =e^{{\cal G}
       \tau}c(\tau,\tau^{\prime})$ into (\ref{A7}) we obtain
       
       \begin{equation}
       c(\tau,\tau^{\prime})=\int_{0}^{\tau}e^{-{\cal G}
       \tau^{\prime\prime}}
       \frac{\partial}{\partial\tau^{\prime\prime}}
       \delta(\tau^{\prime\prime}
       -\tau^{\prime})\;d\tau^{\prime\prime}+c(\tau^{\prime}).
       \end{equation}
       
       \noindent In this way 
       
       \begin{equation}
       \label{A8}
       \Sigma(\tau^{\prime\prime},\tau^{\prime})
       =e^{{\cal G}
       \tau^{\prime\prime}}[c(\tau^{\prime})
       -\delta(\tau^{\prime})]
       +\delta(\tau^{\prime\prime}-\tau^{\prime})
       +{\cal G}\theta
       (\tau^{\prime\prime}-\tau^{\prime})e^{{\cal G}
       (\tau^{\prime\prime}
       -\tau^{\prime})}
       \end{equation}
       
       \noindent and using the initial condition(\ref{A6}),
        taking into account
        $\Sigma(0,\tau^{\prime})=c(\tau^{\prime})$, we obtain
        
        \begin{equation}
        \label{A9}
        c(\tau^{\prime})-\delta(\tau^{\prime})=-\frac{{\cal G}
        e^{{\cal G}}
        e^{-{\cal G}\tau^{\prime}}}{1+e^{{\cal G}}}.
        \end{equation}
        
        \noindent Substituting (\ref{A9}) into (\ref{A8})
         we obtain
        
        \begin{equation}
        \label{A10}
        \Sigma(\tau,\tau^{\prime})=\delta(\tau-\tau^{\prime})+
        \frac{{\cal G}}{2}e^{{\cal G}(\tau-\tau^{\prime})}
        \left[\epsilon(\tau
        -\tau^{\prime})-\tanh\frac{{\cal G}}{2}\right].
        \end{equation}
        
        \noindent Inserting (\ref{A10}) into (\ref{A11})
         results in
        
        \begin{eqnarray}
        G(\tau,\tau^{\prime})&=&\frac{{\cal G}}
        {4}\int e^{{\cal G}(\tau-
        \lambda^{\prime})}\left[\epsilon(\tau-\lambda^{\prime})
        -\tanh\frac
        {{\cal G}}{2}\right]\epsilon(\lambda^{\prime}
        -\tau^{\prime})\;
        d\lambda^{\prime}\nonumber\\
        &=&\frac{1}{2}e^{{\cal G}(\tau-\tau^{\prime})}
        \left[\epsilon(\tau-\tau^{\prime})
        -\tanh\frac{{\cal G}}{2}\right].
        \end{eqnarray}
        
        \noindent We observe that if you take the partial
         derivative in 
        relation to $\tau$ of the expression (\ref{A12})
         the inverse matrix is 
        given by
        
        \begin{eqnarray}
        {\tilde \Lambda}^{-1}(\tau,\tau^{\prime})
        &=&\frac{1}{2}\frac
        {\partial\Sigma}{\partial\tau}\\
        &=&\frac{1}{2}\frac{\partial}{\partial\tau}
        \delta(\tau-\tau^{\prime})
        +\frac{{\cal G}^2}{4}e^{{\cal G}
        (\tau-\tau^{\prime})}\left[
        \epsilon(\tau-\tau^{\prime})-\tanh\frac{{\cal G}}
        {2}\right]\nonumber\\
        &&+\frac{{\cal G}}{2}e^{{\cal G}(\tau-\tau^{\prime})}
        \delta(\tau-\tau^{\prime}),\nonumber
        \end{eqnarray}
        
        \noindent and the partial derivative in relation
         to $\tau^{\prime}$ 
        of the expression (\ref{A11}) gives us
        
        \begin{equation}
        \Sigma(\tau,\tau^{\prime})=-\frac{\partial G}
        {\partial\tau^{\prime}}.
        \end{equation}                                                                               
      

\begin{thebibliography}{99}
      \bibitem{R1} D. M. Gitman, Sh. M. Shvartsman
       and W. da Cruz, 
      Braz. J. Phys. 
      {\bf 24} (1994) 844; Preprint IFUSP/P1053,
       June 1993
      \bibitem{R2} D. M. Gitman and Sh. M. Shvartsman,
       Phys. Lett {\bf B318} 
      (1993) 122; Phys. Lett {\bf B331} (1994) 449
      \bibitem{R3} A. M. Polyakov, Mod. Phys. Lett {\bf A3}
       (1988) 325; in 
      Proc. Les Houches Summer School, {\bf Vol XLIX}
       305 eds. E. Br\'ezin
       and J. Zinn-Justin 
       ( North Holland, Amsterdam, 1990)
      \bibitem{R4} D. M. Gitman, S. I. Zlatev
       and W. da Cruz, Braz. J. Phys. 
      {\bf 26} (1996) 419
      \bibitem{R5} E. S. Fradkin and D. M.
       Gitman, Phys. Rev, {\bf 44} (1991) 
      3220
      \bibitem{R6} D. M. Gitman and
       A. V. Saa, Class. Quant. Grav. {\bf 10} 
      (1993) 1447
      \bibitem{R7} J. Schwinger, Phys. Rev.
       {\bf 82} (1951) 664
      \end{thebibliography}
      \end{document}